\documentclass[reprint,superscriptaddress,amsmath,amssymb,aps]{revtex4-2}
\usepackage{graphicx}
\usepackage{dcolumn}
\usepackage{bm}
\usepackage{dcolumn}
\usepackage{braket}
\usepackage{color}

\begin{document}

\title{Nonreciprocal Phonon Propagation in a Metallic Chiral Magnet}
\author{T. Nomura}
\email{tnomura@mail.dendai.ac.jp}
\affiliation{Institute for Solid State Physics, University of Tokyo, Kashiwa, Chiba 277-8581, Japan}
\affiliation{Tokyo Denki University, Adachi, Tokyo 120-8551, Japan}
\author{X.-X. Zhang}
\email{xiaoxiao.zhang@riken.jp}
\affiliation{RIKEN Center for Emergent Matter Science (CEMS), Wako 351-0198, Japan}
\author{R. Takagi} 
\affiliation{Department of Applied Physics, The University of Tokyo, Tokyo 113-8656, Japan}
\affiliation{PRESTO, Japan Science and Technology Agency (JST), Kawaguchi 332-0012, Japan}
\author{K. Karube}
\affiliation{RIKEN Center for Emergent Matter Science (CEMS), Wako 351-0198, Japan}
\author{A. Kikkawa}
\affiliation{RIKEN Center for Emergent Matter Science (CEMS), Wako 351-0198, Japan}
\author{Y. Taguchi}
\affiliation{RIKEN Center for Emergent Matter Science (CEMS), Wako 351-0198, Japan}
\author{Y. Tokura}
\affiliation{RIKEN Center for Emergent Matter Science (CEMS), Wako 351-0198, Japan}
\affiliation{Department of Applied Physics, The University of Tokyo, Tokyo 113-8656, Japan}
\affiliation{Tokyo College, University of Tokyo, Tokyo 113-8656, Japan}
\author{S. Zherlitsyn}
\affiliation{Hochfeld-Magnetlabor Dresden (HLD-EMFL), Helmholtz-Zentrum Dresden-Rossendorf, 01328 Dresden, Germany}
\author{Y. Kohama}
\affiliation{Institute for Solid State Physics, University of Tokyo, Kashiwa, Chiba 277-8581, Japan}
\author{S. Seki} 
\affiliation{Department of Applied Physics, The University of Tokyo, Tokyo 113-8656, Japan}
\affiliation{PRESTO, Japan Science and Technology Agency (JST), Kawaguchi 332-0012, Japan}

\date{\today}

\begin{abstract}
The phonon magnetochiral effect (MChE) is the nonreciprocal acoustic and thermal transports of phonons caused by the simultaneous breaking of the mirror and time-reversal symmetries.
So far, the phonon MChE has been observed only in a ferrimagnetic insulator Cu$_2$OSeO$_3$, where the nonreciprocal response disappears above the Curie temperature of 58 K.
Here, we study the nonreciprocal acoustic properties of a room-temperature ferromagnet Co$_9$Zn$_9$Mn$_2$ for unveiling the phonon MChE close to the  room temperature.
Surprisingly, the nonreciprocity in this metallic compound is enhanced at higher temperatures and observed up to 250 K.
This clear contrast between insulating Cu$_2$OSeO$_3$ and metallic Co$_9$Zn$_9$Mn$_2$ suggests that metallic magnets have a mechanism to enhance the nonreciprocity at higher temperatures.
From the ultrasound and microwave-spectroscopy experiments, we conclude that the magnitude of the phonon MChE of Co$_9$Zn$_9$Mn$_2$ mostly depends on the Gilbert damping, which increases at low temperatures and hinders the magnon-phonon hybridization.
Our results suggest that the phonon nonreciprocity could be further enhanced by engineering the magnon band of materials.
\end{abstract}
\maketitle

When a chiral material is placed in a magnetic field, nonreciprocal properties arise from the simultaneous breaking of the mirror and time-reversal symmetries.
Such nonreciprocal properties are due to magnetochiral effect (MChE) \cite{Tokura18rev,Atzori21rev} observed for various (quasi)particle propagations, including photons \cite{Rikken97,Vallet01,Koerdt03,Train08,Okamura15,Atzori21}, electrons \cite{Rikken01,Krstic02,Pop14,Yokouchi17,Aoki19,Kitaori21}, magnons \cite{Iguchi15,Seki16,Takagi17,Seki20,Ogawa21}, and phonons \cite{Nomura19}.
For the case of phonons, the sound velocity in a chiral material becomes different for parallel and antiparallel propagations with respect to the magnetic field $\bf H$ \cite{Nomura19}. 
Because of the symmetry origin, the MChE is expected for any chiral materials although the microscopic mechanism and magnitude depend on the system.
Since the nonreciprocal properties are closely related to functionalities, the MChE is an attractive mechanism for novel devices (ex. single-phase diodes and circulators).

\begin{figure}[tb]
\centering
\includegraphics[width=8.2cm]{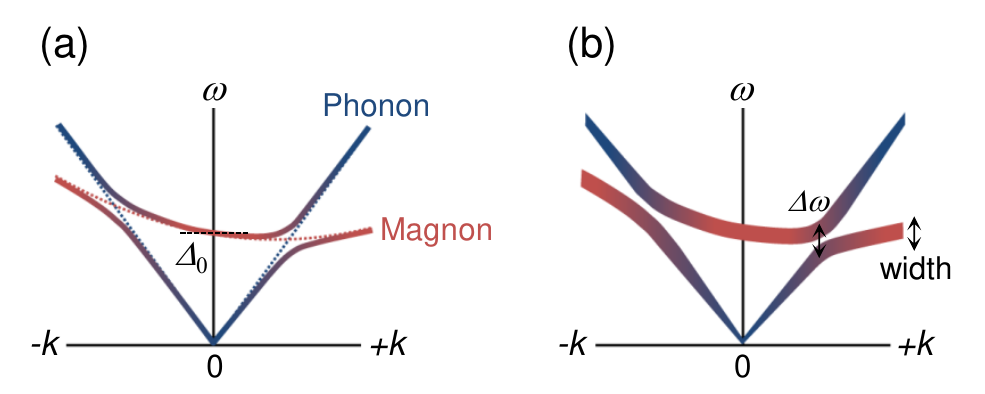}
\caption{\label{fig:hybr}
Phonon MChE caused by the magnon-phonon hybridization.
(a) Dispersion relations of the magnon and acoustic-phonon bands near the $\Gamma$ point under magnetic fields ${\bf H}||{\bf k}$.
(b) Effect of the magnon-dispersion broadening.
Note that only one of the circularly polarized phonon mode hybridizes with magnons, since only right-handed polarization exists for ferromagnetic-type spin waves with respect to the magnetization.
The ultrasound frequency in this study was always much smaller than the magnon gap $\Delta_0/2\pi$.
}
\end{figure}

So far, the phonon MChE has been reported only for the ferrimagnetic insulator Cu$_2$OSeO$_3$ \cite{Nomura19}.
The phonon MChE is explained by a magnon-phonon hybridization \cite{Nomura19,Tereshchenko18}.
Because of the Dzyaloshinskii-Moriya (DM) interaction, the magnon dispersion in Cu$_2$OSeO$_3$ is asymmetric for wave vector $\bf k$ parallel and antiparallel to $\bf H$ \cite{Seki16,Kataoka87}.
When the magnon-phonon hybridization is allowed by symmetry, the phonon dispersion is asymmetrically deformed by the band repulsion, leading to the nonreciprocal sound velocity for $\pm \bf k$ [Fig.~\ref{fig:hybr}(a)].
In this case, the MChE is observed only below the Curie temperature $T_\mathrm{C}\sim 58$ K \cite{Bos08,Seki12Science}.
For realizing the MChE at room temperature, experimental exploration on higher-$T_\mathrm{C}$ magnets is important.
Moreover, the investigation of metallic magnets allows for exploring the phonon nonreciprocity in presence of conduction electrons.

Our target material Co$_9$Zn$_9$Mn$_2$ has a $\beta$-Mn-type structure, which belongs to the chiral space group $P4_132$ ($P4_332$) \cite{Hori07,Xie13}.
The magnetic properties of the series of (Co$_{0.5}$Zn$_{0.5}$)$_{20-x}$Mn$_x$ alloys have been systematically studied, and various exciting properties related to their chiral magnetism have been reported \cite{Takagi17,Tokunaga15,Karube16,Karube17,Morikawa17,Yu18,Karube18,Bocarsly19,Nagase19,Karube20,Nagase21,Preissinger21,Peng21,Shimojima21}.
Among this series, Co$_9$Zn$_9$Mn$_2$ has a high Curie temperature $T_\mathrm{C}\sim 400$ K.
Co$_9$Zn$_9$Mn$_2$ has a helimagnetic ground state, while other Mn-rich compounds become spin glass at low temperature due to the geometrical frustration \cite{Karube18}.
The spin-glass state could complicate the analysis of the magnon dispersion and its temperature dependence.
In addition, the magnon MChE and the magnon dispersion of Co$_9$Zn$_9$Mn$_2$ have already been studied \cite{Takagi17}.
Therefore, Co$_9$Zn$_9$Mn$_2$ is a promising system to explore the room-temperature phonon MChE.

Single crystals of Co$_9$Zn$_9$Mn$_2$ were grown by the Bridgman method as described in our previous papers \cite{Karube16,Karube17,Karube18}.
For the ultrasound measurements, we used a right-handed single crystal ($P4_132$) with the sample length of 1.9 mm along the [110] axis.
We employed the ultrasound pulse-echo technique with a phase-sensitive detection for the sound-velocity measurement \cite{Zherlitsyn14}.
In this study, we investigated three acoustic modes $c_\mathrm{L}=(c_{11}+c_{12}+2c_{44})/2$ (${\bf k}|| {\bf u}||[110]$, $v=4.0$ km/s), $c_\mathrm{T}=(c_{11}-c_{12})/2$  (${\bf k}|| [110]$, ${\bf u}||[1\overline{1}0]$, $v=2.1$ km/s),  and $c_{44}$ (${\bf k}|| [110]$, ${\bf u}||[001]$, $v=2.4$ km/s), where $\bf k$ and $\bf u$ are the propagation and displacement vectors, respectively.
We used the ultrasound frequency up to 530 MHz by high-harmonic generation of the LiNbO$_3$ resonance transducers.
At higher frequency, the acoustic attenuation became too strong to analyze the nonreciprocity.
In this study, the experiments were performed in the Faraday geometry with ${\bf H}|| {\bf k} || [110]$.
In addition, we performed the microwave absorption spectroscopy on a crystal from the same batch.
The microwave absorption ($\Delta S_{11}$) caused by magnetic resonance was monitored by using a coplanar waveguide and a vector-network analyzer.
For details, see Refs.~\cite{Takagi17,Takagi21}.

\begin{figure}[tb]
\centering
\includegraphics[width=8.4cm]{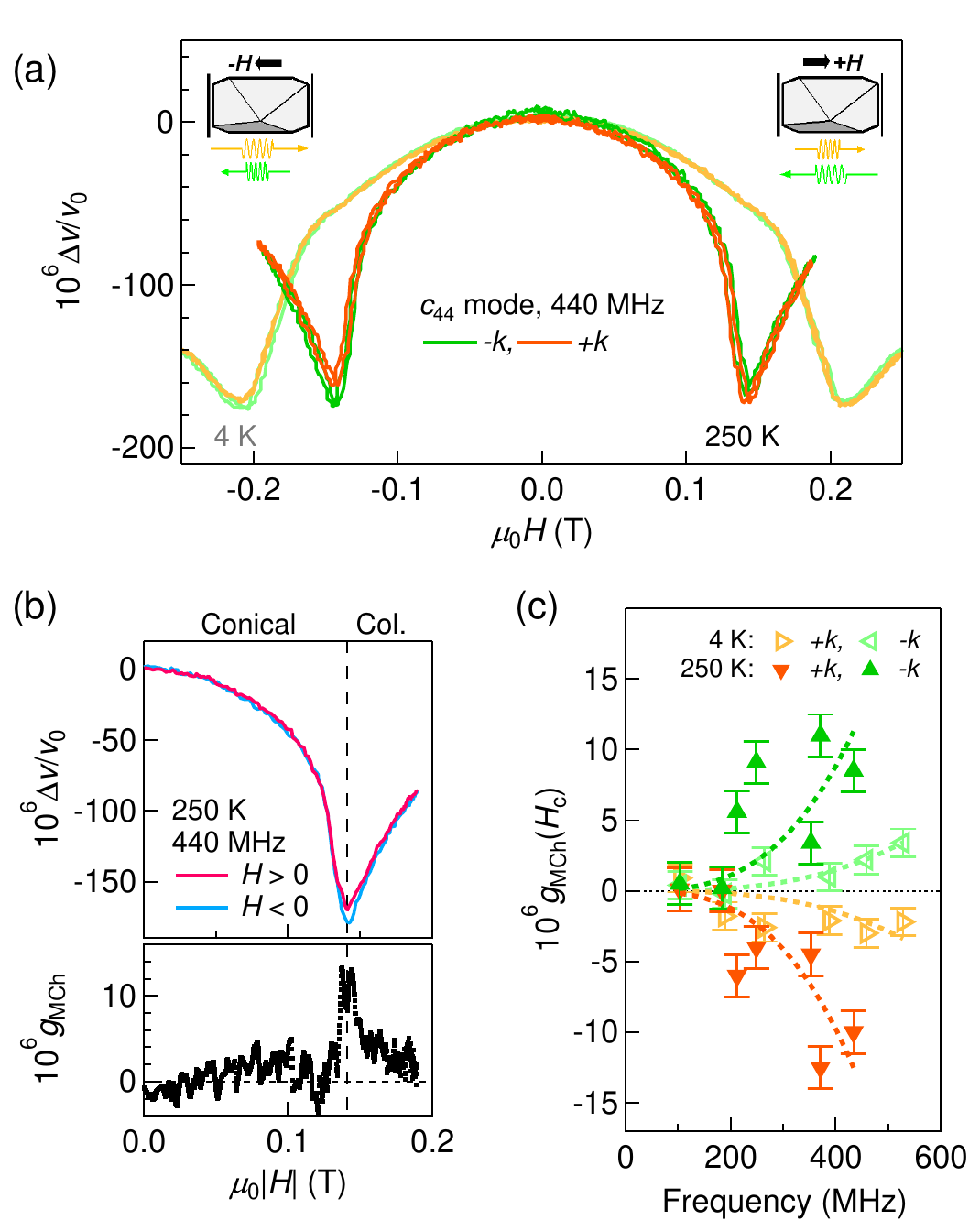}
\caption{\label{fig:phonon}
Phonon MChE in Co$_9$Zn$_9$Mn$_2$.
(a) Relative change of the sound velocity as a function of magnetic field, $\Delta v/v_0(H)$, of the $c_{44}$ acoustic mode at the ultrasound frequency of 440 MHz.
The results at 250 K (4 K) are shown by the orange and green (light orange and light green) curves.
Schematics of the experimental configuration and the nonreciprocity are shown.
(b) $\Delta v/v_0(H)$ and the magnitude of the magnetochiral effect $g_\mathrm{MCh}$ as a function of the absolute value of the magnetic field.
The data for $-k$, 440 MHz at 250 K from Fig.~\ref{fig:phonon}(a) are used.
(c) Ultrasound frequency dependence of the magnitude of the magnetochiral effect at the transition field $g_\mathrm{MCh}(H_\mathrm{c})$.
The results for the $c_{44}$ mode at 4 and 250 K are fitted by Eq.~(\ref{eq:signal1}) (dashed lines).
}
\end{figure}

Figure~\ref{fig:phonon}(a) shows the magnetic-field dependence of the relative change of the sound velocity $\Delta v/v_0$ of the $c_{44}$ acoustic mode at 4 and 250~K.
The magnetic transition from a conical to a collinear state is observed at $H_\mathrm{c}= 0.21$~T (4~K) and 0.14~T (250~K).
The sound velocities for $-\bf k$ and $+\bf k$ at $H_\mathrm{c}$ are slightly different, indicating the nonreciprocal sound propagation.
The sign of the difference becomes opposite for $-H_\mathrm{c}$ and $+H_\mathrm{c}$, which is the characteristic feature of the phonon MChE.
We note that the other acoustic modes ($c_\mathrm{L}$ and $c_\mathrm{T}$) show weaker $\Delta v/v_0$ and nonreciprocal responses, indicating the weaker magnetoelastic coupling.
These results are presented in the Supplemental Material (SM) \cite{Supple}.

\begin{figure}[tb]
\centering
\includegraphics[width=8.6cm]{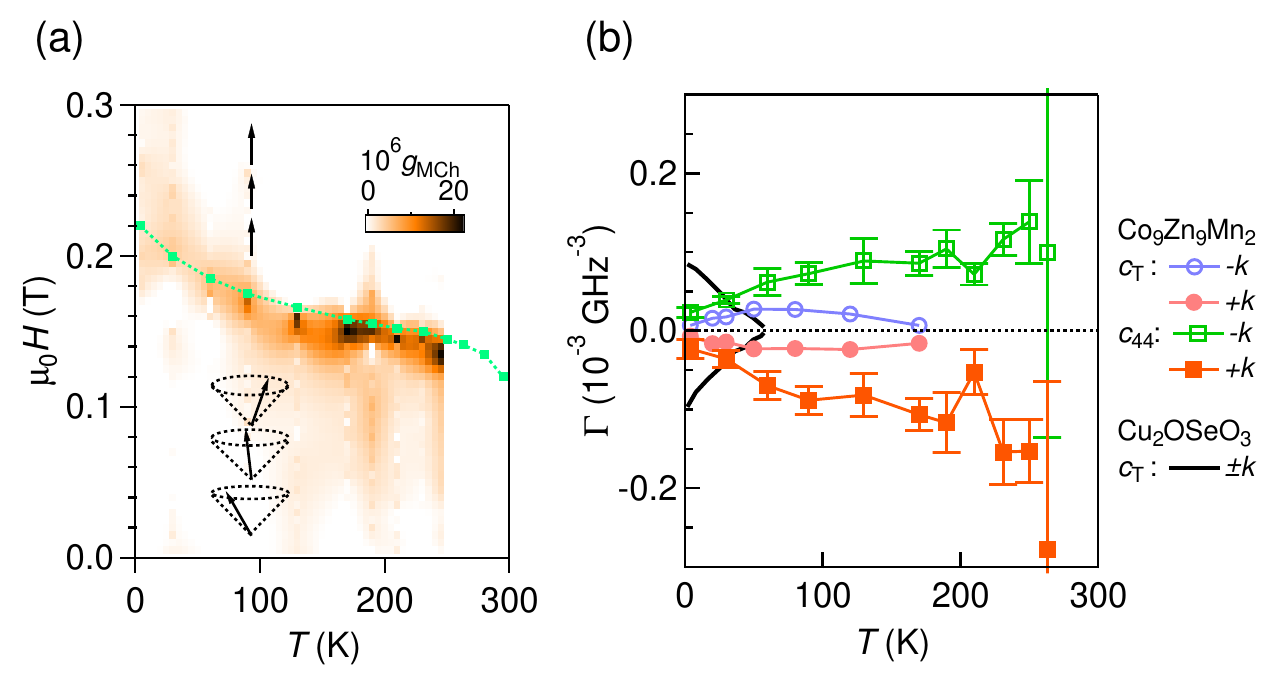}
\caption{\label{fig:color}
(a) Contour plot of $g_\mathrm{MCh}$ mapped on the $T$-$H$ phase diagram. 
The conical-collinear phase boundary (green curve) is determined by the anomalies in $\Delta v/v_0$. 
Here, the results at 440 MHz are used.
(b) Temperature dependence of the magnitude of the phonon MChE.
The coefficient $\Gamma$ is obtained by fitting the frequency dependence of $g_\mathrm{MCh}(H_\mathrm{c})$ [Fig.~\ref{fig:phonon}(c)].
The results of Cu$_2$OSeO$_3$ \cite{Nomura19} are shown for comparison.
}
\end{figure}

Figure \ref{fig:phonon}(b) plots the results for $-\bf k$ as a function of the absolute value of the magnetic field.
The magnitude of the phonon magnetocihral effect $g_\mathrm{MCh}$ defined as \cite{Nomura19}
\begin{equation}
g_\mathrm{MCh}(|{\bf H}|)=\frac{\Delta v(+{\bf H})}{v_0}-\frac{\Delta v(-{\bf H})}{v_0}=\frac{v(+{\bf H})-v(-{\bf H})}{v_0}
\label{eq:gmch}
\end{equation}
is plotted in the lower panel.
The nonreciprocal response takes maximum at $H_\mathrm{c}$, and then rapidly decreases when the field is further increased.
The nonreciprocal magnitude at the transition field $g_\mathrm{MCh}(H_\mathrm{c})$ as a function of the ultrasound frequency is plotted in Fig.~\ref{fig:phonon}(c).
The nonreciprocity nonlinearly enhances at higher frequencies.
$g_\mathrm{MCh}(H_\mathrm{c})$ at 250 K is larger than that at 4~K.
The contour plot of $g_\mathrm{MCh}$ at 440~MHz is mapped on the $T$-$H$ phase diagram [Fig.~\ref{fig:color}(a)].

The experimental results show that the magnetochiral effect of Co$_9$Zn$_9$Mn$_2$ is enhanced at $H_\mathrm{c}$ and at higher ultrasound frequency.
Similar features are observed in Cu$_2$OSeO$_3$ and explained by the magnon-phonon hybridization mechanism \cite{Nomura19}.
The magnon dispersion of Co$_9$Zn$_9$Mn$_2$ is also asymmetric for $\pm \bf k$ because of the DM interaction \cite{Takagi17}.
When such magnon band hybridizes with a phonon band, the anticrossing asymmetrically deforms the phonon band [Fig.~\ref{fig:hybr}(a)], leading to the nonreciprocal acoustic properties.
At the conical-collinear transition, the magnon gap becomes small, and the hybridization occurs close to the $\Gamma$ point (Fig.~\ref{fig:hybr}).
However, the magnon frequency is more than one order of magnitude higher than the ultrasound frequency in this study, and only the effect due to the band repulsion is observed in our experiments.
Here, we emphasize that the electron-phonon hybridization cannot explain the characteristic enhancement of the phonon MChE at $H_\mathrm{c}$ and the field dependence in the collinear phase.
The energy scale of the magnetic field (0.3 T) is too small to change the electronic band structure of metals.
We, thus, conclude that the rapid decrease of the phonon MChE above $H_\mathrm{c}$ is related to the magnon band gap which also rapidly opens above $H_\mathrm{c}$.

The magnon-phonon hybridization is mediated by the DM interactions or magnetic anisotropy modulated by shear strains \cite{Nomura19}.
Since the DM interactions in chiral magnets are strongly modulated by shear strains \cite{Koretsune,Shibata}, we treat the former as the leading term. 
The magnitude of the MChE is expressed as \cite{Nomura19},
\begin{equation}\label{eq:signal1}
g_\mathrm{MCh} = \frac{ 4\gamma^2 \braket{S^z}^2 S^2 D^3 k^3} {c \Delta_0^2} = \Gamma f^3.
\end{equation}
Here, $\gamma,S,D,c,\Delta_0$ are the magnetoelastic coupling constant, the total spin moment, the DM-interaction coefficient, the elastic constant, and the magnon gap, respectively.
Since the wave vector $k$ is proportional to the ultrasound frequency $f$, the equation is rewritten as the empirical form with the phonon-MChE coefficient $\Gamma$.

Based on Eq. (\ref{eq:signal1}), the frequency dependence of $g_\mathrm{MCh}(H_\mathrm{c})$ [Fig. \ref{fig:phonon}(c)] is fitted and the temperature dependence of $\Gamma$ is obtained as Fig.~\ref{fig:color}(b).
The MChE is observed up to 170 K and 250 K for the $c_\mathrm{T}$ and $c_{44}$ modes, respectively.
$|\Gamma|$ is always larger for the $c_{44}$ than that for the $c_\mathrm{T}$ mode, which again indicates the stronger magnetoelastic coupling of the $c_{44}$ mode.
$|\Gamma|$ of the $c_{44}$ mode increases towards higher temperatures and suddenly becomes undetectable at 260 K.
This is because of the sudden decrease of $\Delta v/v_0(H_\mathrm{c})$ and increase of the acoustic attenuation around this temperature \cite{Supple}.
This characteristic temperature might be due to the decreased anisotropy at high temperatures \cite{Preissinger21}.
Another extrinsic origin might be the glass transition of the bond connecting the transducers and the sample. 
Above the glass transition temperature, the transverse sound propagation is strongly attenuated.
In this case, the thermal transport experiments might be an alternative technique to detect the phonon MChE \cite{Hirokane20}.
However, the simultaneous decrease of the acoustic anomaly $\Delta v/v_0(H_\mathrm{c})$ and the nonreciprocity $g_\mathrm{MCh}(H_\mathrm{c})$ (Fig. S1 in SM \cite{Supple}) indicates that the observations come from the intrinsic effect.
We note that the bonding conditions usually only affect the amplitude of the transmitted sound and does not affect the obtained $\Delta v/v_0$.

For a quantitative comparison, the results of Cu$_2$OSeO$_3$ [$c_\mathrm{T}=(c_{11}-c_{12})/2$ mode] \cite{Nomura19} are plotted by the black lines.
The maximum value of $\Gamma$ in Co$_9$Zn$_9$Mn$_2$ is larger than that in Cu$_2$OSeO$_3$ by the factor of $\sim$1.5.
$|\Gamma|$ is tending to increase with temperature in Co$_9$Zn$_9$Mn$_2$, though it is decreasing in Cu$_2$OSeO$_3$ [Fig. \ref{fig:color}(b)]. 
In the following, we discuss the reason of the different temperature dependence by comparing the case of insulator (Cu$_2$OSeO$_3$) and metal (Co$_9$Zn$_9$Mn$_2$).

The negative-$T$ coefficient of $|\Gamma|$ for Cu$_2$OSeO$_3$ is due to the temperature dependence of the ordered moment $\braket{S^z}$ \cite{Nomura19}.
Near the magnetic phase transition, $\braket{S^z}$ scales as $\sqrt{|T-T_\mathrm{C}|}$, leading the the $T$-linear dependence of $g_\mathrm{MCh}$ [Eq. (\ref{eq:signal1})].
Similarly to the case of Co$_9$Zn$_9$Mn$_2$, the temperature dependence of $\braket{S^z}$ almost scales as $\sqrt{|T-T_\mathrm{C}|}$ \cite{Bocarsly19}, however, it only results in the negative-$T$ coefficient of $|\Gamma|$.
Therefore, the positive-$T$ coefficient of $|\Gamma|$ (high-temperature enhancement of the nonreciprocity) originates from another property, which is strongly dependent on temperature and overcomes the contribution from $\braket{S^z}$.

\begin{figure}[tb]
\centering
\includegraphics[width=8.6cm]{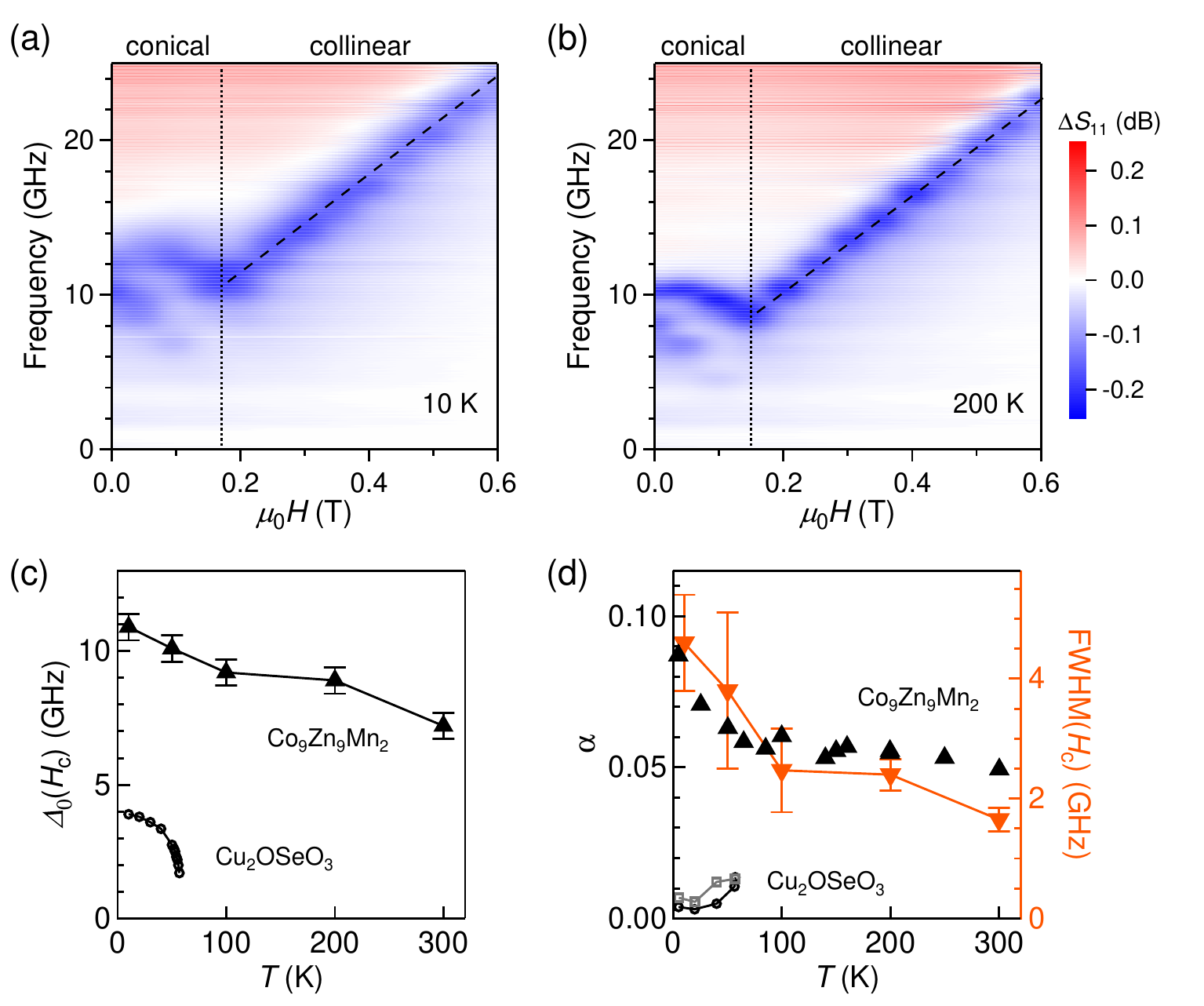}
\caption{\label{fig:da}
Magnetic-field dependence of the microwave-absorption spectra ($\Delta S_{11}$) at (a) 10 K and (b) 200 K. 
The ferromagnetic resonance lines are shown by the dashed lines for guide. 
The conical-collinear phase boundary is shown by the dotted line.
(c) Temperature dependence of the magnon-gap frequency at the transition field.
(d) Temperature dependence of the Gilbert damping coefficient $\alpha$ (left) and the FWHM at the transition field (right).
The results for Cu$_2$OSeO$_3$ are taken from Ref.~\cite{Stasinopoulos17} and shown for comparison.
Square (circle) symbols represent the result for ${\bf H}||[111]$ (${\bf H}||[100]$).
$\alpha$ of Co$_9$Zn$_9$Mn$_2$ is taken from Ref.~\cite{Preissinger21}.
}
\end{figure}

Another $T$-dependent parameter in Eq. (\ref{eq:signal1}) is the magnon gap $\Delta_0$.
Figures \ref{fig:da}(a) and \ref{fig:da}(b) show the microwave-absorption spectra as a function of the magnetic field at 10~K and 200~K, respectively.
With this technique, the obtained spectra reflect the magnon gap at $k=0$.
The overall features at 10~K and 200~K are similar.
In the conical phase, two branches of spin waves soften towards $H_\mathrm{C}$.
In the collinear phase, single absorption peak linearly increases as a function of the magnetic field (dashed line).
Slight temperature dependence is observed in the magnon gap and width of the absorption peak.
Figure \ref{fig:da}(c) plots the temperature dependence of the magnon gap at the transition field $\Delta_0(H_\mathrm{c})$ determined from the crossing point of the dashed line and the dotted line (the conical-collinear boundary).
For comparison, the results of Cu$_2$OSeO$_3$ are also shown \cite{Stasinopoulos17}.
The gap of Co$_9$Zn$_9$Mn$_2$ increases towards lower temperatures by the factor of $\sim1.2$, which would lead to the decrease of $g_\mathrm{MCh}(H_\mathrm{c})$ by the factor of 1.4 [Eq.~(2)].
This parameter cannot account for the observed factor of $\Gamma(\mathrm{200\ K})/\Gamma(\mathrm{10\ K}) \sim10$ [Fig. \ref{fig:color}(b)].

\begin{table*}
\caption{\label{tab:table1} Summary of the magnetic and elastic parameters. The unit cells of Co$_9$Zn$_9$Mn$_2$ and Cu$_2$OSeO$_3$ are taken per Co ion and per 4Cu cluster, respectively. The last column presents the calculated coefficient of the phonon MChE based on Eq.~(2).}
\begin{ruledtabular}
\begin{tabular}{lccccccc|c}
 & $S$ & $a_0$ & $D/a_0^2$ & $\Delta_0(H_\mathrm{c})/2\pi$ & FWHM$(H_\mathrm{c})$ &  $c$ && $g_\mathrm{MCh}/\gamma^2 k^3$\\
  &  & (\AA) & (J m$^{-2}$) & (GHz) & (GHz) &  (GPa) && (m$^3$)\\
\hline
Co$_9$Zn$_9$Mn$_2$ (200 K) \cite{Takagi17,Karube17} & 0.75 & 3.1 & 1.2$\times 10^{-3}$ & 8.9 & 2.4 & $c_{44}=45$ && 43$\times$10$^{-30}$\\
Co$_9$Zn$_9$Mn$_2$ (10 K)  & & & & 10.9 & 4.6 & $c_{44}=47$ && \\
Cu$_2$OSeO$_3$ (30 K) \cite{Seki16} & 0.44 & 4.5 & 3.4$\times 10^{-4}$ & 3.7 & 0.4 & $c_\mathrm{T}=27$ && 11$\times$10$^{-30}$\\
\end{tabular}
\end{ruledtabular}
\end{table*}

Next, the temperature dependence of the full width at half maximum (FWHM) of the magnetic resonance at $H_\mathrm{c}$ is plotted in Fig. \ref{fig:da}(d) right axis with orange triangles.
The broadening of the absorption peak is observed below 100~K.
The width of the magnon absorption ($\Delta f$) is related to the Gilbert damping parameter $\alpha$ as \cite{Yu12,Stasinopoulos17}
\begin{equation}\label{eq:Gilbert}
\Delta f = 2 \alpha f_\mathrm{r} + \Delta f_0,
\end{equation}
where $f_\mathrm{r}$ is the resonance frequency and $\Delta f_0$ is the extrinsic broadening.
The Gilbert damping parameters of Co$_9$Zn$_9$Mn$_2$ \cite{Preissinger21} and Cu$_2$OSeO$_3$ \cite{Stasinopoulos17} are plotted in Fig.~\ref{fig:da}(d) left axis.
The temperature dependence of $\alpha$ and the width of the magnon absorption is in a proportional relation as suggested by Eq.(3).
The magnon damping of Co$_9$Zn$_9$Mn$_2$ increases towards lower temperatures.
This is a typical behavior for magnetic alloys, where magnons are scattered by conduction electrons \cite{Yu12,Gilmore10,Faehnle11}.
At lower temperature, the elecrton scattering time increases and the larger momentum transfer at the intraband relaxation results in the enhanced damping.
For the case of Co$_9$Zn$_9$Mn$_2$, disorder of Mn spins at low temperatures might also be the reason of the increased $\alpha$ \cite{Preissinger21}.
On the other hand, in the case of magnetic insulators, the magnon-phonon and magnon-magnon scatterings are suppressed at lower temperatures (less populated magnons and phonons), leading to the smaller $\alpha$.
Therefore, the opposite temperature dependence of $\alpha$ is a clear contrast between magnetic insulators and metals, which is a key to understand the temperature dependence of the phonon MChE of Co$_9$Zn$_9$Mn$_2$.

A hybridized state of a magnon and a phonon is called a magnon polaron \cite{Kikkawa,Flebus,Kittel}.
When the magnon scattering time $\tau$ is too short to well form the hybridized state (namely $\alpha$ is too large), the anticrossing gap $\Delta \omega$ is effectively reduced.
For instance, when $1/\tau \gg \Delta\omega$, the magnon-polaron bands become more smeared and short-lived in this weak-coupling case \cite{Kikkawa}. 
Substituting $\tau=1/\alpha \omega$, this condition reads $\alpha \gg \Delta \omega/\omega$.
In other words, the magnon polaron cannot be formed when the magnon line width is too large compared to the anticrossing gap [Fig. \ref{fig:hybr}(b)].
The anticrossing gap is estimated as (see SM for details \cite{Supple}),
\begin{equation}
\begin{split}
\Delta\omega=\frac{\gamma\braket{S^z}\sqrt{S}D}{a_0^2 \sqrt{\rho s}v_0^2}{\omega^*}^\frac{3}{2},
\end{split}
\end{equation}
where $a_0$, $\rho$, $s=\hbar/V_0$, and $\omega^*$ are the lattice constant, the mass density, the reduced Planck constant normalized by the unit-cell volume, and the angular frequency at the hybridization point.
Substituting the parameters of Co$_9$Zn$_9$Mn$_2$, $\Delta\omega / 2\pi$ is estimated to be of the order of 10 MHz, using $\gamma \sim 10$.
The estimated $\Delta \omega/\omega$ is of the order of $10^{-3}$, which is consistent with $\Delta v/v_0 \sim 10^{-4}$ because one can roughly use $\omega = vk$ for small $k$.
Therefore, the damping factor $\alpha \sim 0.1$ is much larger than $\Delta \omega/\omega$, indicating that the magnon-phonon hybridization occurs in a weak coupling regime.
In this case, the magnon linewidth (and the magnon scattering time) can considerably affect the hybridization.
Indeed, the anomaly at the phase transition $\Delta v(H_c)/v_0$ decreases below 100 K where $\alpha$ also increases (Fig. S1(b) \cite{Supple}), indicating the weaker magnon-phonon hybridization.
Such an effectively weaker magnon-phonon coupling accordingly leads to smaller anticrossing gap and also reduces nonreciprocity $g_\mathrm{MCh}$. 
Quantitative relation of how $\alpha$ affects $g_\mathrm{MCh}$ exceeds the present theory based on sharp magnon-polaron bands and is left for future investigation.

The above discussion suggests that the phonon MChE can be further enhanced by reducing the magnon linewidth, which is related to the impurity or defect in the crystal.
Particularly, for the case of (Co$_{0.5}$Zn$_{0.5}$)$_{20-x}$Mn$_x$ alloys, the end material Co$_{10}$Zn$_{10}$ has relatively small $\alpha$ \cite{Preissinger21} which might be advantageous to realize a larger nonreciprocity.
In this respect, magnetic insulators typically have smaller $\alpha$ than metals where the magnon-electron scattering is inevitable.
In fact, even for the case of Cu$_2$OSeO$_3$, which shows rather small $\alpha$ comparable to yttrium gallium garnet \cite{Stasinopoulos17}, the strong-coupling condition $\alpha \ll \Delta \omega/\omega$ is not satisfied.
This indicates that the magnon linewidth is an important factor even for discussing the phonon MChE in magnetic insulators.

Nevertheless, the phonon-MChE coefficient $\Gamma$ of Co$_9$Zn$_9$Mn$_2$ is 1.5 times larger than that of Cu$_2$OSeO$_3$.
This is because of the strong DM interaction in Co$_9$Zn$_9$Mn$_2$.
By using the parameters summarized in Table I and Eq. (2), in fact, the expected $g_\mathrm{MChE}$ for Co$_9$Zn$_9$Mn$_2$ is four times larger than Cu$_2$OSeO$_3$, assuming similar $\gamma$.
Therefore, this system has a great potential to realize larger phonon MChE at room temperature, if the magnon linewidth could be controlled.

In conclusion, we demonstrate the phonon MChE, the nonreciprocal sound propagation in the chiral-lattice magnetic metal Co$_9$Zn$_9$Mn$_2$ up to 250 K.
In contrast to the case in Cu$_2$OSeO$_3$, the phonon MChE of Co$_9$Zn$_9$Mn$_2$ is enhanced at higher temperatures.
This temperature dependence is due to the magnon linewidth.
The magnon linewidth of Co$_9$Zn$_9$Mn$_2$ rapidly increases below 100 K and results in the weak coupling of the magnon-phonon hybridization, leading to the decreased gap and nonreciprocity.
Controlling the magnon dispersion could further enhance the nonreciprocal properties of these chiral crystals.

\vspace{1cm}
We thank N. Nagaosa for fruitful discussions.
We acknowledge the support of the HLD at HZDR, member of the European Magnetic Field Laboratory (EMFL), and the DFG through the Collaborative Research Center SFB 1143 (project-id 247310070).
This work was partly supported by JSPS KAKENHI, Grant-in-Aid for Scientific Research (JP19K23421, JP20K14403, JP20H00349, JP20K15164, JP21H04440, JP21H04990, JP21K13876, JP21K18595, JP22H04965), JSPS Bilateral Joint Research Projects (JPJSBP120193507),  JST PRESTO (JPMJPR20B4),  JST CREST (JPMJCR1874, JPMJCR20T), and Katsu Research Encouragement Award of the University of Tokyo, Asahi Glass Foundation and Murata Science Foundation.


\begin{thebibliography}{99}
\bibitem{Tokura18rev}
Y. Tokura and N. Nagaosa, Nat. Commun. {\bf9}, 3740 (2018).
\bibitem{Atzori21rev}
M. Atzori, C. Train, E. A. Hillard, N. Avarvari, and G. L. J. A. Rikken, Chirality {\bf33}, 844--857 (2021).
\bibitem{Rikken97}
G. L. J. A. Rikken and E. Raupach, Nature (London) {\bf390}, 493-494 (1997).
\bibitem{Vallet01}
M. Vallet, R. Ghosh, A. Le Floch, T. Ruchon, F. Bretenaker, and J. Y. Th\'epot, Phys. Rev. Lett. {\bf87}, 183003 (2001).
\bibitem{Koerdt03}
C. Koerdt, G. D\"uchs, and G. L. J. A. Rikken, Phys. Rev. Lett. {\bf91}, 073902 (2003).
\bibitem{Train08}
C. Train, R. Gheorghe, V. Krstic, L. M. Chamoreau, N. S. Ovanesyan, G. L. J. A. Rikken, M. Gruselle, and M. Verdaguer, Nat. Mater. {\bf7}, 729-734 (2008).
\bibitem{Okamura15}
Y. Okamura, F. Kagawa, S. Seki, M. Kubota, M. Kawasaki, and Y. Tokura, Phys. Rev. Lett. {\bf114}, 197202 (2015).
\bibitem{Atzori21}
M. Atzori, H. D. Ludowieg, \'A. Valent\'in-P\'erez, M. Cortijo, I. Breslavetz, K. Paillot, P. Rosa, C. Train, J. Autschbach, E. A. Hillard, and G. L. J. A. Rikken, Sci. Adv. {\bf7}, eabg2859 (2021).
\bibitem{Rikken01}
G. L. J. A. Rikken, J. F\"olling, and P. Wyder, Phys. Rev. Lett. {\bf87}, 236602 (2001).
\bibitem{Krstic02}
V. Krstic, S. Roth, M. Burghard, K. Kern, and G. L. J. A. Rikken, J. Chem. Phys. {\bf117}, 11315-11319 (2002).
\bibitem{Pop14}
F. Pop, P. Auban-Senzier, E. Canadell, G. L. J. A. Rikken, and N. Avarvari, Nat. Commun. {\bf5}, 3757 (2014).
\bibitem{Yokouchi17}
T. Yokouchi, N. Kanazawa, A. Kikkawa, D. Morikawa, K. Shibata, T. Arima, Y. Taguchi, F. Kagawa, and Y. Tokura, Nat. Commun. {\bf8}, 4757 (2017).
\bibitem{Aoki19}
R. Aoki, Y. Kousaka, and Y. Togawa, Phys. Rev. Lett. {\bf122}, 057206 (2019).
\bibitem{Kitaori21}
A. Kitaori, N. Kanazawa, H. Ishizuka, T. Yokouchi, N. Nagaosa, and Y. Tokura, Phys. Rev. B {\bf103}, L220410 (2021).
\bibitem{Seki16}
S. Seki, Y. Okamura, K. Kondou, K. Shibata, M. Kubota, R. Takagi, F. Kagawa, M. Kawasaki, G. Tatara, Y. Otani, and Y. Tokura, Phys. Rev. B {\bf93}, 235131 (2016).
\bibitem{Takagi17}
R. Takagi, D. Morikawa, K. Karube, N. Kanazawa, K. Shibata, G. Tatara, Y. Tokunaga, T. Arima, Y. Taguchi, Y. Tokura, and S. Seki, Phys. Rev. B {\bf95}, 220406(R) (2017).
\bibitem{Iguchi15}
Y. Iguchi, S. Uemura, K. Ueno, and Y. Onose, Phys. Rev. B {\bf92}, 184419 (2015).
\bibitem{Seki20}
S. Seki, M. Garst, J. Waizner, R. Takagi, N. D. Khanh, Y. Okamura, K. Kondou, F. Kagawa, Y. Otani, and Y. Tokura, Nat. Commun. {\bf11}, 256 (2020).
\bibitem{Ogawa21}
N. Ogawa, L. K\"ohler, M. Garst, S. Toyoda, S. Seki, and Y. Tokura, Proc. Natl. Acad. Sci. U.S.A. {\bf118}, e2022927118 (2021).
\bibitem{Nomura19}
T. Nomura, X.-X. Zhang, S. Zherlitsyn, J. Wosnitza, Y. Tokura, N. Nagaosa, and S. Seki, Phys. Rev. Lett. {\bf122}, 145901 (2019).
\bibitem{Tereshchenko18}
A. A. Tereshchenko, A. S. Ovchinnikov, I. Proskurin, E. V. Sinitsyn, and J. Kishine, Phys. Rev. B {\bf97}, 184303 (2018).
\bibitem{Kataoka87}
M. Kataoka, J. Phys. Soc. Jpn. {\bf56}, 3635-3647 (1987).
\bibitem{Bos08}
J.-W. G. Bos, C. V. Colin, and T. T. M. Palstra, Phys. Rev. B {\bf78}, 094416 (2008).
\bibitem{Seki12Science}
S. Seki, X. Z. Yu, S. Ishiwata, and Y. Tokura, Science {\bf336}, 198-201 (2012).
\bibitem{Hori07}
T. Hori, H. Shiraisha, and Y. Ishii, J. Magn. Magn. Mater. {\bf310}, 1820--1822 (2007).
\bibitem{Xie13}
W. Xie, S. Thimmaiah, J. Lamsal, J. Liu, T. W. Heitmann, D. Quirinale, A. I. Goldman, V. Pecharsky, and G. J. Millera, Inorg. Chem. {\bf52}, 9399–9408 (2013).
\bibitem{Tokunaga15}
Y. Tokunaga, X. Z. Yu, J. S. White, H. M. R\o{}nnow, D. Morikawa, Y. Taguchi, and Y. Tokura, Nat. Commun. {\bf6}, 7638--7644 (2015).
\bibitem{Karube16}
K. Karube, J. S. White, N. Reynolds, J. L. Gavilano, H. Oike, A. Kikkawa, F. Kagawa, Y. Tokunaga, H. M. R\o{}nnow, Y. Tokura, and Y. Taguchi, Nat. Mater. {\bf15}, 1237--1243 (2016).
\bibitem{Karube17}
K. Karube, J. S. White, D. Morikawa, M. Bartkowiak, A. Kikkawa, Y. Tokunaga, T. Arima, H. M. R\o{}nnow, Y. Tokura, and Y. Taguchi, Phys. Rev. Mater. {\bf1}, 074405 (2017).
\bibitem{Morikawa17}
D. Morikawa, X. Yu, K. Karube, Y. Tokunaga, Y. Taguchi, T. Arima, and Y. Tokura, Nano Lett. {\bf17}, 1637--1641 (2017).
\bibitem{Yu18}
X. Z. Yu, W. Koshibae, Y. Tokunaga, K. Shibata, Y. Taguchi, N. Nagaosa, and Y. Tokura, Nature (London) {\bf564}, 95 (2018).
\bibitem{Karube18}
K. Karube, J. S. White, D. Morikawa, C. D. Dewhurst, R. Cubitt, A. Kikkawa, X. Yu, Y. Tokunaga, T. Arima, H. M. R\o{}nnow, Y. Tokura, and Y. Taguchi, Sci. Adv. {\bf4}: eaar7043 (2018).
\bibitem{Bocarsly19}
J. D. Bocarsly, C. Heikes, C. M. Brown, S. D. Wilson, and R. Seshadri, Phys. Rev. Materials {\bf3}, 014402 (2019).
\bibitem{Nagase19}
T. Nagase, M. Komatsu, Y. G. So, T. Ishida, H. Yoshida, Y. Kawaguchi, Y. Tanaka, K. Saitoh, N. Ikarashi, M. Kuwahara, and M. Nagao, Phys. Rev. Lett. {\bf123},137203 (2019).
\bibitem{Karube20}
K. Karube, J. S. White, V. Ukleev, C. D. Dewhurst, R. Cubitt, A. Kikkawa, Y. Tokunaga, H. M. R\o{}nnow, Y. Tokura,  and Y. Taguchi, Phys. Rev. B {\bf102}, 064408 (2020).
\bibitem{Nagase21}
T. Nagase, Y. G. So, H. Yasui, T. Ishida, H. K. Yoshida, Y. Tanaka, K. Saitoh, N. Ikarashi, Y. Kawaguchi, M. Kuwahara, and M. Nagao, Nat. Commun. {\bf12}, 3490 (2021).
\bibitem{Preissinger21}
M. Prei\ss inger, K. Karube, D. Ehlers, B. Szigeti, H.-A. Krug von Nidda, J. S. White, V. Ukleev, H. M. R\o{}nnow, Y. Tokunaga, A. Kikkawa, Y. Tokura, Y. Taguchi, and I. K\'ezsm\'arki, npj Quantum Materials {\bf6}:65 (2021).
\bibitem{Peng21}
L. C. Peng, K. Karube, Y. Taguchi, N. Nagaosa, Y. Tokura, and X. Z. Yu, Nat. Commun. {\bf12}, 6797 (2021).
\bibitem{Shimojima21}
T. Shimojima, A. Nakamura, X. Z. Yu, K. Karube, Y. Taguchi, Y. Tokura, and K. Ishizaka, Sci. Adv. 7, eabg1322 (2021).
\bibitem{Zherlitsyn14}
S. Zherlitsyn, S. Yasin, J. Wosnitza, A. A. Zvyagin, A. V. Andreev, and V. Tsurkan, Low Temp. Phys. {\bf40}, 123-133 (2014).
\bibitem{Takagi21}
R. Takagi, M. Garst, J. Sahliger, C. H. Back, Y. Tokura, and S. Seki, Phys. Rev. B {\bf104}, 144410 (2021).
\bibitem{Supple}
See Supplemental Material at http:// for the experimental details and theoretical derivation of the anticrossing gap at the magnon-phonon hybridization.
\bibitem{Koretsune}
T. Koretsune, N. Nagaosa, and R. Arita, Sci. Rep. {\bf5}, 13302 (2015). 
\bibitem{Shibata}
K. Shibata {\it et al.}, Nat. Nanotech. {\bf10}, 589-592 (2015).
\bibitem{Hirokane20}
Y. Hirokane, Y. Nii, H. Masuda, and Y. Onose, Sci. Adv. {\bf6}, eabd3703 (2020).
\bibitem{Stasinopoulos17}
I. Stasinopoulos, S. Weichselbaumer, A. Bauer, J. Waizner, H. Berger, S. Maendl, M. Garst, C. Pfleiderer, and D. Grundler, Appl. Phys. Lett. {\bf111}, 032408 (2017).
\bibitem{Yu12}
H. Yu, R. Huber, T. Schwarze, F. Brandl, T. Rapp, P. Berberich, G. Duerr, and D. Grundler, Appl. Phys. Lett. {\bf100}, 262412 (2012).
\bibitem{Gilmore10}
K. Gilmore, M. D. Stiles, J. Seib, D. Steiauf, and M. F\"ahnle, Phys. Rev. B {\bf81}, 174414 (2010).
\bibitem{Faehnle11}
M. F\"ahnle and C. Illg, J. Phys.: Condens. Matter {\bf23}, 493201 (2011).
\bibitem{Kikkawa}
T. Kikkawa, K. Shen, B. Flebus, R. A. Duine, K. Uchida, Z. Qiu, G. E. W. Bauer, and E. Saitoh, Phys. Rev. Lett. {\bf117}, 207203 (2016).
\bibitem{Flebus}
B. Flebus, K. Shen, T. Kikkawa, K. Uchida, Z. Qiu, E. Saitoh, R. A. Duine, and G. E. W. Bauer, Phys. Rev. B {\bf95}, 144420 (2017).
\bibitem{Kittel}
C. Kittel, Phys. Rev. {\bf110}, 836--841 (1958).
\end{thebibliography}
\end{document}